# Machine Learning Models for the Identification of Cardiovascular Diseases Using UK Biobank Data

Sheikh Mohammed Shariful Islam[1], Moloud Abrar[2], Teketo Tegegne[1], Liliana Loranjo[3], Chandan Karmakar[2], Md Abdul Awal[4], Md. Shahadat Hossain[5], Muhammad Ashad Kabir[6], Mufti Mahmud[7], Abbas Khosravi[2], George Siopis[8], Jeban C Moses[1], Ralph Maddison[1]

1 Institute for Physical Activity and Nutrition, Deakin University, Melbourne, Australia
2 School of Information Technology, Deakin University, Melbourne, Australia
3 Westmead Applied Research Centre, The University of Sydney, Sydney, Australia
4 School of Electrical Engineering and Computer Science, The University of Queensland, Brisbane, Australia
5 School of Computer, Data, and Mathematical Sciences, Western Sydney University, Sydney, Australia
6 School of Computing, Mathematics and Engineering, Charles Sturt University, Bathurst, Australia
7 Department of Computer Science, Nottingham Trent University, Nottingham, UK
8 Faculty of Medicine, Health and Human Sciences, Macquarie University, Sydney, Australia

**Corresponding author:** shariful.islam@deakin.edu.au and akabir@csu.edu.au

**Author emails:** shariful.islam@deakin.edu.au, m.abdar@deakin.edu.au, teketo.tegegne@deakin.edu.au, liliana.laranjo@sydney.edu.au, karmakar@deakin.edu.au, m.awal@uq.edu.au, shahadat_qs@iubat.edu, akabir@csu.edu.au, mufti.mahmud@ntu.ac.uk, abbas.khosravi@deakin.edu.au, georgios.siopis@students.mq.edu.au, jcmoses@deakin.edu.au, ralph.maddison@deakin.edu.au

**Keywords:** Machine Learning, Heart Diseases, Primary Prevention, Artificial Intelligence, Prediction.

**ABSTRACT**

**Background**: Machine learning models have the potential to identify cardiovascular diseases (CVDs) early and accurately in primary healthcare settings, which is crucial for delivering timely treatment and management. Although population-based CVD risk models have been used traditionally, these models often do not consider variations in lifestyles, socioeconomic conditions, or genetic predispositions. Therefore, we aimed to develop machine learning models for CVD detection using primary healthcare data, compare the performance of different models, and identify the best models.

**Methods**: We used data from the UK Biobank study, which included over 500,000 middle-aged participants from different primary healthcare centers in the UK. Data collected at baseline (2006--2010) and during imaging visits after 2014 were used in this study. Baseline characteristics, including sex, age, and the Townsend Deprivation Index, were included. Participants were classified as having CVD if they reported at least one of the following conditions: heart attack, angina, stroke, or high blood pressure. Cardiac imaging data such as electrocardiogram and echocardiography data, including left ventricular size and function, cardiac output, and stroke volume, were also used. We used 9 machine learning models (LSVM, RBFSVM, GP, DT, RF, NN, AdaBoost, NB, and QDA), which are explainable and easily interpretable. We reported the accuracy, precision, recall, and F-1 scores; confusion matrices; and area under the curve (AUC) curves.

**Results**: RBFSVM, GP, DT, and AdaBoost were the best performing models (accuracy: 96), and LSVM showed the weakest prediction performance (accuracy: 0.69). While RBFSVM and GP could significantly classify people with CVD better, LSVM and AdaBoost could classify healthy participants more accurately. The LSVM has a better false positive rate but a worse false negative rate. Moreover, the RBFSVM and GP have better performance in terms of false negatives. RBFSVM, GP, DT, NN, and AdaBoost produced the best AUC values (accuracy: 0.96) for CVD detection.

**Conclusions**: Our study shows that machine learning models can detect CVD via primary healthcare data. RBFSVM, GP, and AdaBoost were the best performing methods. However, the RBFSVM significantly outperforms the other models in terms of false negatives and is recommended for detecting CVD. Future studies are recommended to validate our findings in different population groups.

**INTRODUCTION**

Cardiovascular diseases (CVDs) are the leading cause of death worldwide [1-3]. In 2019, an estimated 18.6 million people died from CVD [4]. Among these deaths, 49.2% were due to ischemic heart disease [4]. In the United Kingdom (UK), 7.6 million people live with CVD [5], and 255 per 100,000 people died from CVD in 2019 [6]. The 2019 UK's 1.18 million hospital admissions were due to CVD, with an estimated cost of £19 billion [6]. Most CVDs can be prevented by addressing lifestyle risk behaviors, such as smoking, alcohol consumption, unhealthy diet and physical inactivity [7, 8].

CVD is diagnosed via a range of laboratory tests and imaging studies. However, challenges remain in diagnosing and managing CVD [9]. If CVD is not diagnosed early or if the diagnosis is not accurate, patients will not receive the needed advice, treatment, and management [9]. For example, 20–40% of heart attacks are reported in patients with previously undiagnosed CVD, highlighting a missed opportunity for early intervention [10]. Most people have reported experiencing at least one diagnostic error in their lifetime, which could have devastating consequences [11]. Approximately 10% of deaths and 6–17% of hospital adverse events in the United States are attributed to diagnostic errors [11]. Early detection of CVD is crucial for delivering timely treatment to patients and minimizing CVD progression [12]. Population-based CVD risk models are invaluable in estimating whether a person is at low, intermediate, or high risk of having a CVD event in the next 10 years [13-15]. However, the currently available risk models do not perform well for every person, as they do not consider variations in lifestyles, socioeconomic conditions, and genetic predispositions [16].

In recent years, machine learning (ML) algorithms have been widely used for different disease detection methods. Automated systems using ML that can detect and localize CVD at an early stage have great potential for CVD prevention and management [17]. However, the choice of classification algorithms is complex, and there is often a trade-off between model accuracy and transparency. The application of deep neural networks in medicine has been widely used, but the results are often unexplainable and presented as a 'black box', making it difficult for clinicians to comprehend. Deep learning requires large datasets where the features are sorted in descending order and the most important features concerning the target variables receive the topmost position. For example, a deep learning technique such as the conventional neural network employed for CVD detection requires passing through multiple hidden layers of the neural network. As a consequence, a considerable number of nondescriptive features are generated that cannot be categorized or visualized via contemporary feature importance techniques. These rationales inspire the preference for traditional ML models instead of deep learning models [18]. Furthermore, while machine learning models cannot mine the temporal information of data compared with deep learning models [19], they have good performance, ease of use, and low computational burden. There is a dearth of studies using ML to detect CVD via large primary care datasets [20]. In this study, we aimed to use different ML models for CVD detection from a large dataset to identify the best performing models. The results from this study will facilitate comparisons between different ML models for use in clinical settings for accurate and timely diagnosis of CVD.

**METHODS**

**Datasets**

In this study, we used data from the United Kingdom (UK) Biobank study [21]. The UK Biobank study collected data throughout the UK from middle-aged participants aged more than 40 years [21, 22]. More than half a million (501,490) participants were enrolled during the initial assessment visit (2006--2010) when data on their sociodemographic conditions, baseline health status, and medical conditions diagnosed by a physician [22] were recorded. During a subsequent visit in 2014 or later, 39,619 participants provided imaging data. For this study, we only used data collected at baseline (2006--2010) and during 2014 and later.

**Variables and definitions**

The baseline characteristics included sociodemographic information, including sex, age, and the Townsend Deprivation Index (TSDI). The TSDI, a census-based deprivation index introduced by Townsend [23], comprises four census variables: households without a car, overcrowded households, households not owner-occupied, and persons unemployed. The TSDI can be constructed for any geographical area where the aforementioned four variables are well available. We included clinical data, such as systolic and diastolic blood pressure, 12-lead electrocardiograph (ECG) measurements of the ventricular rate, P duration, QRS duration, automated ECG diagnoses, number of automated diagnostic comments recorded during 12-lead ECG, suspicious flag for 12-lead ECG, left ventricular (LV) size and function, LV ejection fraction, LV end-diastolic volume, LV end-systolic volume, LV stroke volume, cardiac output, cardiac index, average heart rate, body surface area and vascular/heart problems diagnosed by a physician. The cardiac imaging data were obtained after the first subsequent imaging visit, starting in 2014. For this study, we used an ECG at rest and left ventricular size and function. A participant was classified as having CVD if they were diagnosed by a physician; reported having a heart attack, angina, stroke, or high blood pressure; or had vascular/heart problems detected via cardiac imaging data.

**Data analysis**

We performed data preprocessing, including multiple imputations for missing data management in the Python programming language, via sci-kit-learn and pandas. We divided the data into training (60%) and test (40%) sets. Using the training dataset, we applied the nine most commonly used ML models for identifying CVD, including linear support vector machine (LSVM), radial basis function SVM (RBFSVM), Gaussian process (GP), decision tree (DT), random forest (RF), neural network (NN), AdaBoost, naïve

Bayes (NB), and quadratic discriminant analysis (QDA). We used the test dataset to evaluate model performance via different model performance metrics, such as precision, recall, F1 score, accuracy, and area under the curve (AUC) [1-5][24-28]. These ML models and evaluation metrics have been widely used in the literature for classification tasks [6-8][29-31]. The dataset was balanced, and there were no differences between the Micro and weighted averages.

**RESULTS**

Nine different traditional ML algorithms, namely, LSVM, RBFSVM, GP, DT, RF, NN, AdaBoost, NB, and QDA, were used for the classification task. Table 1 presents the precision, recall, F1 score, and accuracy results of the different models. Figure 2 presents the confusion matrices for the models, and Figure 3 presents their area under the curve (AUC).

Table 1. Comparison of nine machine learning models in terms of precision, recall, F1 score, and accuracy. LSVM=Lagrangian support vector machine; RBFSVM=Radial basis function support vector machine; GP=Gaussian processes; DT=Decision trees; RF= Random forest; NN=Neural network; NB=Naïve Bayes; QDA=Quadratic discriminant analysis

| Method | Class | Precision | Recall | F1-score | Accuracy |
|---|---|---|---|---|---|
| LSVM | 0 Healthy | 0.62 | 0.98 | 0.76 | 0.69 |
| | 1 People with CVD | 0.95 | 0.40 | 0.56 | |
| | Weighted average | 0.79 | 0.69 | 0.66 | |
| RBFSVM | 0 | 1.00 | 0.92 | 0.96 | **0.96** |
| | 1 | 0.92 | 1.00 | 0.96 | |
| | Weighted average | 0.96 | 0.96 | 0.96 | |
| | 0 | 0.99 | 0.92 | 0.96 | |
| GP | 1 | 0.93 | 0.99 | 0.96 | |
| | Weighted average | 0.96 | 0.96 | 0.96 | **0.96** |
| DT | 0 | 0.99 | 0.92 | 0.96 | |
| | 1 | 0.93 | 0.99 | 0.96 | |
| | Weighted average | 0.96 | 0.96 | 0.96 | **0.96** |
| RF | 0 | 0.98 | 0.92 | 0.95 | |
| | 1 | 0.93 | 0.98 | 0.95 | |
| | Weighted average | 0.95 | 0.95 | 0.95 | 0.95 |
| NN | 0 | 0.99 | 0.92 | 0.95 | |
| | 1 | 0.93 | 0.99 | 0.96 | |
| | Weighted average | 0.96 | 0.96 | 0.95 | 0.95 |
| AdaBoost | 0 | 0.99 | 0.93 | 0.96 | |
| | 1 | 0.93 | 0.99 | 0.96 | |
| | Weighted average | 0.96 | 0.96 | 0.96 | **0.96** |
| NB | 0 | 0.95 | 0.92 | 0.94 | |
| | 1 | 0.92 | 0.95 | 0.94 | |
| | Weighted average | 0.94 | 0.94 | 0.94 | 0.94 |
| QDA | 0 | 0.99 | 0.92 | 0.95 | |
| | 1 | 0.93 | 0.99 | 0.96 | |
| | Weighted average | 0.96 | 0.95 | 0.95 | 0.95 |

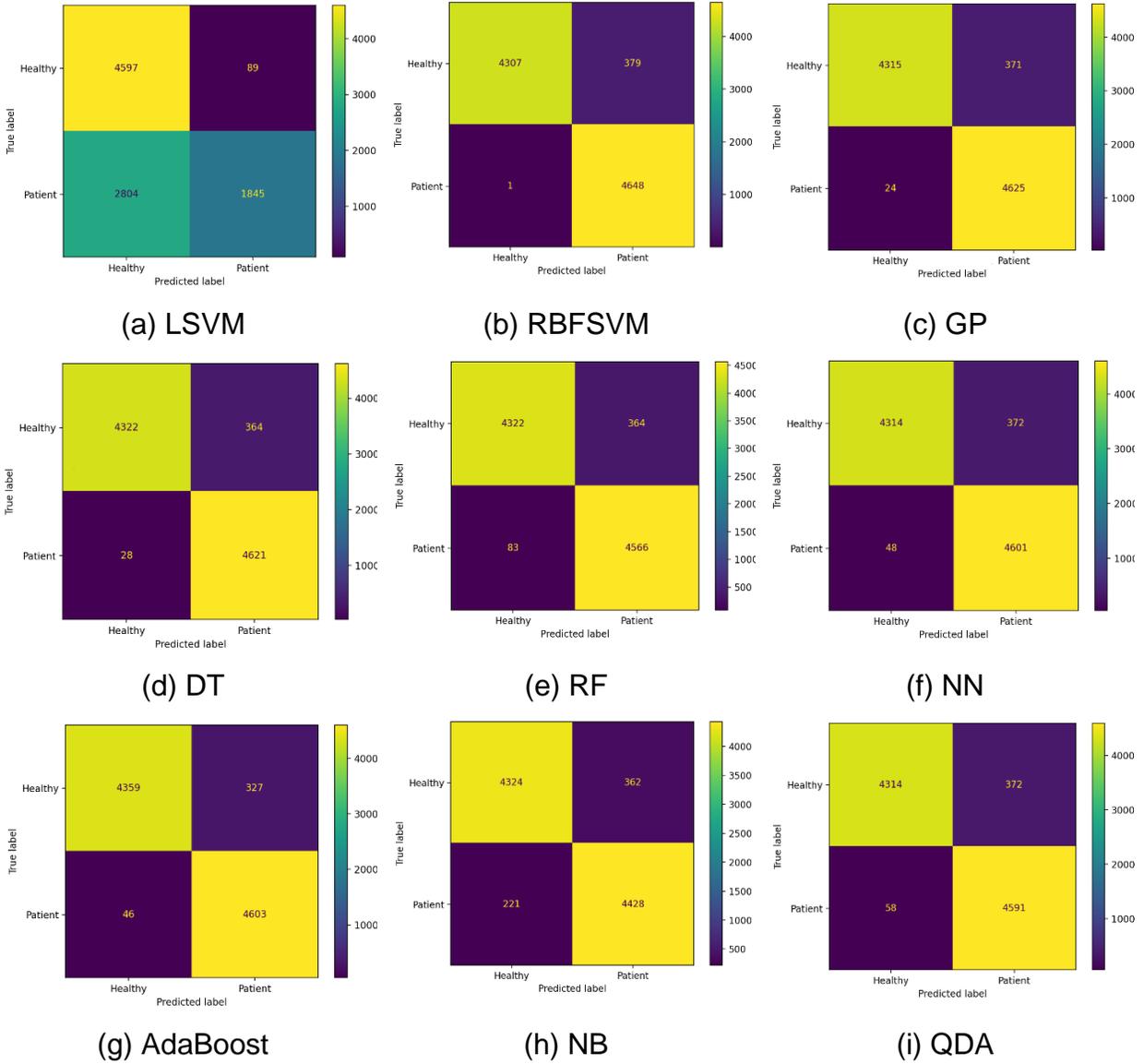

Figure 2. Confusion matrix for applied machine learning algorithms. LSVM=Lagrangian support vector machine; RBFSVM=Radial basis function support vector machine; GP=Gaussian processes; DT=Decision trees; RF= Random forest; NN=Neural network; NB=Naïve Bayes; QDA=Quadratic discriminant analysis

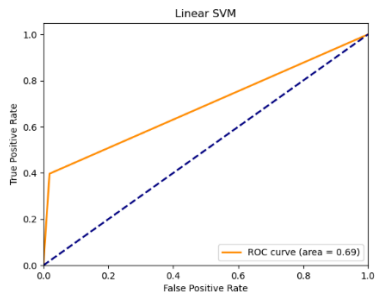 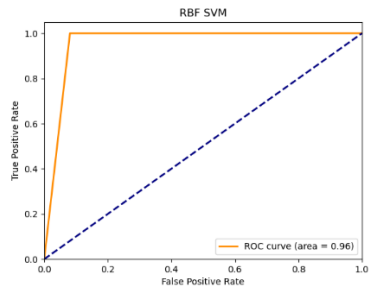 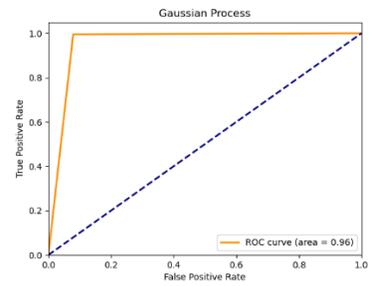

(a) LSVM   (b) RBFSVM   (c) GP

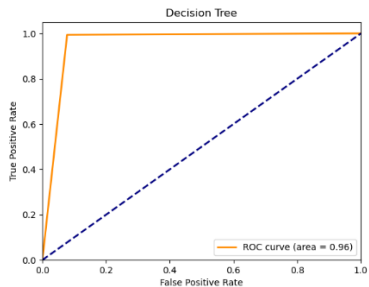 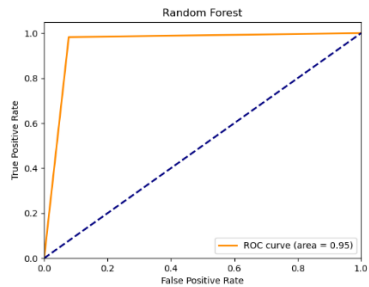 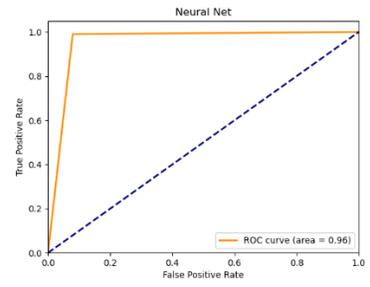

(d) DT   (e) RF   (f) NN

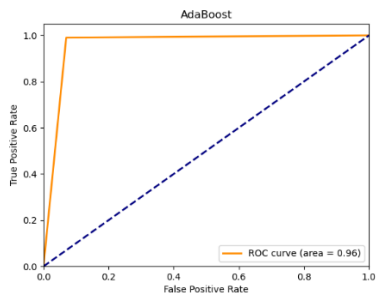 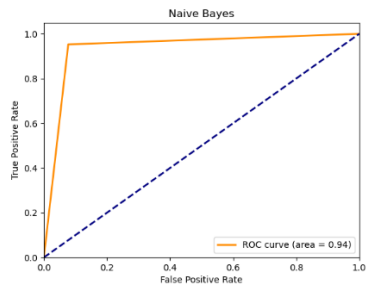 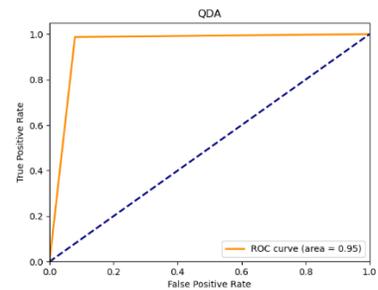

(g) AdaBoost   (h) NB   (i) QDA

Figure 3. AUC curves for the applied machine learning algorithms. LSVM=Lagrangian support vector machine; RBFSVM=Radial basis function support vector machine; GP=Gaussian processes; DT=Decision trees; RF= Random forest; NN=Neural network; NB=Naïve Bayes; QDA=Quadratic discriminant analysis

Table 1 shows that the RBFSVM, GP, DT, and AdaBoost models outperform the other methods. The LSVM (accuracy: 0.69) showed the weakest prediction performance among the methods. Figure 2 shows that RBFSVM (Figure 2 (b)) and GP (Figure 2 (c)) could accurately classify patients with CVD, whereas LSVM (Figure 2 (a)) and AdaBoost (Figure 2(g)) could classify healthy samples much better than the other methods. The LSVM (Figure 2(a)) exhibited a superior false positive rate but also had a worse false negative (FN) rate. Compared with the other applied methods, the RBFSVM (Figure 2(b)) and GP (Figure 2(c)) methods performed better in terms of false negative (FN) rates.

RBFSVM (Figure 3(b)), GP (Figure 3(c)), DT (Figure 3(d)), NN (Figure 3(f)), and AdaBoost (Figure 3(g)) achieved the best AUC values (0.96). The RBFSVM significantly outperforms the other models in terms of the FN rate; therefore, it is selected as the best model in this study.

Supplementary Table 1 compares the results of this study with those of 12 previous studies [8-19] that were conducted from 2019--2022. Three of these studies included participants with CVD [12-14], and nine included participants with CAD [8-11,15-19]. A variety of models were used. Three studies performed better than our study did, reporting accuracies of 100% [17], 99.05% [13], and 97.37% [18], compared with our study's 96% accuracy. Our study performed better than the other nine studies [8-12,14-16,19].

**DISCUSSION**

This study compared nine different ML algorithms for detecting CVD using data from primary healthcare centers in the UK Biobank study. The RBFSVM model exhibited the best performance in terms of the AUC value and false negative rates; therefore, it is our recommended ML algorithm for the detection of CVD in a primary care setting. Our findings have significant implications for the early detection and prevention of CVD. The use of ML algorithms can help identify individuals who are at risk of developing CVD at an early stage and provide timely intervention and management. It also highlights the importance of including socioeconomic and demographic factors in CVD risk prediction models, as these factors can have a significant impact on the development of CVD. Furthermore, our study highlights the potential of explainable machine learning algorithms for CVD risk prediction. By identifying the most important features in predicting CVD, clinicians and researchers can gain a better understanding of the underlying risk factors and develop targeted prevention and management strategies.

Several studies have used ML models to predict CVD risk. Notably, some of the studies listed have used the same datasets (UK Biobank), whereas Li used their own datasets. In addition, we used our own dataset collected from the UK Biobank dataset. A comparison between the studies facilitates the formation of recommendations for guiding clinical practice. Singh et al. [17][32], Ghosh et al. [13][33], and Gupta et al. [18][34] obtained better results than our study did. Although Ghosh et al. [13][33] and Gupta et al. [18][34] achieved better accuracy than our study did, they needed to eliminate some features to reach the goal of achieving better performance. However, Singh et al. [17][32] did not eliminate any features from their datasets. In our study, we trained and validated all the applied ML algorithms with all the features included. We decided to not eliminate any features, despite their different impacts on the accurate diagnosis of disease, to produce a more complete and comprehensive diagnostic model.

The strength of our study is the use of a large primary care dataset with robust and measurable variables. Our study has known limitations. First, we used a single dataset with limited variables and missing data. However, we used multiple imputation techniques to address missing data. Second, we were unable to perform external validation with different datasets. Third, the state-of-the-art classifiers proposed in this study have been applied without optimizing the hyperparameters, model calibration, or feature engineering. Future studies using feature analysis techniques such as SHAP or LIME might improve the functions of the models. Moreover, ensemble learning methods have achieved significant performance in different applications [20-22][35-37]. A new ensemble learning method is recommended.

The expert-in-the-loop scenario is an important approach for closely working between medical experts and AI technologies, which can save experts' time (approximately 80% of their time) in time-intensive tasks. Therefore, proposing an expert-in-the-loop scenario can also be worthy of CVD diagnosis in primary care. Deep learning methods have achieved outstanding performance in different applications. Research comparing traditional machine learning and deep learning methods is warranted. Our study did not consider the uncertainty of the applied methods. Notably, dealing with the uncertainty of a wide variety of machine learning and deep learning methods has become important for obtaining trustworthy predictions [23-27][38-41]. Hence, applying uncertainty quantification methods can be an important future direction.

Our study has significant implications for the early detection and prevention of CVD. The use of machine learning algorithms can help identify individuals who are at risk of developing CVD at an early stage and provide timely intervention and management. Our findings also highlight the importance of including socioeconomic and demographic factors in CVD risk prediction models, as these factors can have a significant impact on the development of CVD. Furthermore, our study highlights the potential of explainable machine learning algorithms for CVD risk prediction. By identifying the most important features in predicting CVD, clinicians and researchers can gain a better understanding of the underlying risk factors and develop targeted prevention and management strategies.

**Conclusion**

In conclusion, our study aimed to identify the potential of machine learning algorithms for the early detection of cardiovascular diseases via data from the UK Biobank study. We found that age, sex, and the Townsend Deprivation Index were among the most important features for predicting CVD. Future studies should explore the potential of machine learning algorithms for CVD risk prediction using data from multiple sources and develop more sophisticated models that can integrate temporal data and other factors.

**Acknowledgments:** We thank the Institute for Physical Activity and Nutrition for subscribing to the UK Biobank data and allowing access. SMS Islam reports support from an Emerging Leadership Fellowship from the National Health and Medical Research Council of Australia (APP1195406) and Vanguard grants from the National Heart Foundation of Australia. SMS Islam has unpaid roles, outside the submitted work, with the IT Committee of the Cardiac Society of Australia and New Zealand, as a volunteer on the Cardiac Devices Committee of the ESC Heart Failure Association, and as a volunteer topic group leader for the WHO-ITU Global Initiative on AI for Health.

**Author contributions:** Conceptualization: SMSI, MA, AK, RM; Data extraction and analysis: TT and MA, First Draft: SMSI, JM, GS, Data interpretation: MAM, MAK, MM, MSH, CK, SMSI; Critical review and comments to improve the manuscript: all authors; Approval to submit: all authors.

**Conflicts of interest:** We have no conflicts of interest to declare.

**Funding:** None.

**Ethics approval and consent to participate:** Not applicable.

**Availability of data and materials**: The data that support the findings of this study are available from the UK Biobank (https://www.ukbiobank.ac.uk/), but restrictions apply to the availability of these data, which were used under license for the current study, and so are not publicly available.

**References**

1. Khan, M. U., Aziz, S., Iqtidar, K., Zaher, G. F., Alghamdi, S., & Gull, M. (2021). A two-stage classification model integrating feature fusion for coronary artery disease detection and classification. *Multimedia Tools and Applications*, 1-30.
1. Vos T, Lim SS, Abbafati C, Abbas KM, Abbasi M, Abbasifard M, Abbasi-Kangevari M, Abbastabar H, Abd-Allah F, Abdelalim A: **Global burden of 369 diseases and injuries in 204 countries and territories, 1990–2019: a systematic analysis for the Global Burden of Disease Study 2019**. *The Lancet* 2020, **396**(10258):1204-1222.
2. Kyu HH, Abate D, Abate KH, Abay SM, Abbafati C, Abbasi N, Abbastabar H, Abd-Allah F, Abdela J, Abdelalim A: **Global, regional, and national disability-adjusted life-years (DALYs) for 359 diseases and injuries and healthy life expectancy (HALE) for 195 countries and territories, 1990–2017: a systematic analysis for the Global Burden of Disease Study 2017. *The Lancet* 2018, **392**(10159):1859-1922.
3. Forouzanfar MH, Afshin A, Alexander LT, Anderson HR, Bhutta ZA, Biryukov S, Brauer M, Burnett R, Cercy K, Charlson FJ: **Global, regional, and national comparative risk assessment of 79 behavioral, environmental and occupational, and metabolic risks or clusters of risks, 1990–2015: a systematic analysis for the Global Burden of Disease Study 2015**. *The lancet* 2016, **388**(10053):1659-1724.
4. Roth GA, Mensah GA, Johnson CO, Addolorato G, Ammirati E, Baddour LM, Barengo NC, Beaton AZ, Benjamin EJ, Benziger CP: **Global burden of cardiovascular diseases and risk factors, 1990–2019: update from the GBD 2019 study**. *Journal of the American College of Cardiology* 2020, **76**(25):2982-3021.
5. **Heart Statistics. BHF Statistics Factsheet - UK** [https://www.bhf.org.uk/what-we-do/our-research/heart-statistics]
6. Cheema KM, Dicks E, Pearson J, Samani NJ: **Long-term trends in the epidemiology of cardiovascular diseases in the UK: insights from the British Heart Foundation statistical compendium**. *Cardiovascular Research* 2022.
7. Lee I-M, Shiroma EJ, Lobelo F, Puska P, Blair SN, Katzmarzyk PT, Group LPASW: **Effect of physical inactivity on major noncommunicable diseases worldwide: an analysis of burden of disease and life expectancy**. *The lancet* 2012, **380**(9838):219-229.
8. World Health Organization: **Global health risks: mortality and burden of disease attributable to selected major risks**: World Health Organization; 2009.
9. McClellan M, Brown N, Califf RM, Warner JJ: **Call to action: urgent challenges in cardiovascular disease: a presidential advisory from the American Heart Association**. *Circulation* 2019, **139**(9):e44-e54.

10. Dalen JE, Alpert JS, Goldberg RJ, Weinstein RS: **The epidemic of the 20th century: coronary heart disease**. *The American journal of medicine* 2014, **127**(9):807-812.
11. Balogh EP, Miller BT, Ball JR: **Improving diagnosis in health care**. 2015.
12. Nagavelli U, Samanta D, Chakraborty P: **Machine Learning Technology-Based Heart Disease Detection Models**. *Journal of Healthcare Engineering* 2022, **2022**.
13. Hippisley-Cox J, Coupland C, Vinogradova Y, Robson J, May M, Brindle P: **Derivation and validation of QRISK, a new cardiovascular disease risk score for the United Kingdom: prospective open cohort study**. *Bmj* 2007, **335**(7611):136.
14. Wilson PW, D'Agostino RB, Levy D, Belanger AM, Silbershatz H, Kannel WB: **Prediction of coronary heart disease using risk factor categories**. *Circulation* 1998, **97**(18):1837-1847.
15. Tunstall-Pedoe H: **Preventing Chronic Diseases. A Vital Investment: WHO Global Report. Geneva: World Health Organization, 2005. pp 200. CHF 30.00. ISBN 92 4 1563001. Also published on http://www. who. int/chp/chronic_disease_report/en**. In.: Oxford University Press; 2006.
16. Hosein A, Stoute V, Chadee S, Singh NR: **Evaluating Cardiovascular Disease (CVD) risk scores for participants with known CVD and non-CVD in a multiracial/ethnic Caribbean sample**. *PeerJ* 2020, **8**:e8232.
17. Khan MA, Algarni F: **A healthcare monitoring system for the diagnosis of heart disease in the IoMT cloud environment using MSSO-ANFIS**. *IEEE Access* 2020, **8**:122259-122269.
18. Elsayad AM, Fakhr M: **Diagnosis of cardiovascular diseases with bayesian classifiers**. *J Comput Sci* 2015, **11**(2):274-282.
19. Lu H, Yao Y, Wang L, Yan J, Tu S, Xie Y, He W: **Research Progress of Machine Learning and Deep Learning in Intelligent Diagnosis of the Coronary Atherosclerotic Heart Disease**. *Computational and Mathematical Methods in Medicine* 2022, **2022**.

20. Weng SF, Reps J, Kai J, Garibaldi JM, Qureshi N: **Can machine-learning improve cardiovascular risk prediction using routine clinical data?** *PloS one* 2017, **12**(4):e0174944.
21. **UK Biobank: a large scale prospective epidemiological resource.** [https://www.ukbiobank.ac.uk/media/gnkeyh2q/study-rationale.pdf]
22. Sudlow C, Gallacher J, Allen N, Beral V, Burton P, Danesh J, Downey P, Elliott P, Green J, Landray M: **UK biobank: an open access resource for identifying the causes of a wide range of complex diseases of middle and old age**. *PLoS medicine* 2015, **12**(3):e1001779.
23. Townsend P: **Deprivation**. *Journal of social policy* 1987, **16**(2):125-146.
24. Abdar M, Salari S, Qahremani S, Lam H-K, Karray F, Hussain S, Khosravi A, Acharya UR, Makarenkov V, Nahavandi S: **UncertaintyFuseNet: robust uncertainty-aware hierarchical feature fusion model with ensemble Monte Carlo dropout for COVID-19 detection**. *Information Fusion* 2022.
25. Abdar M, Wijayaningrum VN, Hussain S, Alizadehsani R, Plawiak P, Acharya UR, Makarenkov V: **IAPSO-AIRS: A novel improved machine learning-based system for wart disease treatment**. *Journal of medical systems* 2019, **43**(7):1-23.
26. Pourpanah F, Shi Y, Lim CP, Hao Q, Tan CJ: **Feature selection based on brain storm optimization for data classification**. *Applied Soft Computing* 2019, **80**:761-775.
27. Pourpanah F, Wang D, Wang R, Lim CP: **A semisupervised learning model based on fuzzy min–max neural networks for data classification**. *Applied Soft Computing* 2021, **112**:107856.
28. Samami M, Akbari E, Abdar M, Plawiak P, Nematzadeh H, Basiri ME, Makarenkov V: **A mixed solution-based high agreement filtering method for class noise detection in binary classification**. *Physica A: Statistical Mechanics and its Applications* 2020, **553**:124219.
29. Abdar M, Acharya UR, Sarrafzadegan N, Makarenkov V: **NE-nu-SVC: a new nested ensemble clinical decision support system for effective diagnosis of coronary artery disease**. *IEEE Access* 2019, **7**:167605-167620.

30. Pourpanah F, Tan CJ, Lim CP, Mohamad-Saleh J: **A Q-learning-based multiagent system for data classification**. *Applied Soft Computing* 2017, **52**:519-531.
31. Roui MB, Zomorodi M, Sarvelayati M, Abdar M, Noori H, Pławiak P, Tadeusiewicz R, Zhou X, Khosravi A, Nahavandi S: **A novel approach based on genetic algorithm to speed up the discovery of classification rules on GPUs**. *Knowledge-Based Systems* 2021, **231**:107419.
32. Singh RS, Gelmecha DJ, Sinha DK: **Expert system based detection and classification of coronary artery disease using ranking methods and nonlinear attributes**. *Multimedia Tools and Applications* 2022:1-28.
33. Ghosh P, Azam S, Jonkman M, Karim A, Shamrat FJM, Ignatious E, Shultana S, Beeravolu AR, De Boer F: **Efficient prediction of cardiovascular disease using machine learning algorithms with relief and LASSO feature selection techniques**. *IEEE Access* 2021, **9**:19304-19326.
34. Gupta A, Kumar R, Arora HS, Raman B: **C-CADZ: computational intelligence system for coronary artery disease detection using Z-Alizadeh Sani dataset**. *Applied Intelligence* 2022, **52**(3):2436-2464.
35. Samami M: **Binary classification of Lupus scientific articles applying deep ensemble model on text data**. In: *2019 Seventh International Conference on Digital Information Processing and Communications (ICDIPC): 2019*: IEEE; 2019: 12-17.
36. Verma P, Awasthi VK, Sahu SK, Shrivas AK: **Coronary Artery Disease Classification Using Deep Neural Network and Ensemble Models Optimized by Particle Swarm Optimization**. *International Journal of Applied Metaheuristic Computing (IJAMC)* 2022, **13**(1):1-25.
37. Yeganeh A, Pourpanah F, Shadman A: **An ANN-based ensemble model for change point estimation in control charts**. *Applied Soft Computing* 2021, **110**:107604.
38. Kompa B, Snoek J, Beam AL: **Second opinion needed: communicating uncertainty in medical machine learning**. *NPJ Digital Medicine* 2021, **4**(1):1-6.

39. Qin Y, Liu Z, Liu C, Li Y, Zeng X, Ye C: **Super-Resolved q-Space deep learning with uncertainty quantification**. *Medical Image Analysis* 2021, **67**:101885.
40. Tanno R, Worrall DE, Kaden E, Ghosh A, Grussu F, Bizzi A, Sotiropoulos SN, Criminisi A, Alexander DC: **Uncertainty modeling in deep learning for safer neuroimage enhancement: demonstration in diffusion MRI**. *NeuroImage* 2021, **225**:117366.
41. Wang Y, Rocková V: **Uncertainty quantification for sparse deep learning**. In: *International Conference on Artificial Intelligence and Statistics: 2020*: PMLR; 2020: 298-308.
42. Abdar M, Książek W, Acharya UR, Tan R-S, Makarenkov V, Pławiak P: **A new machine learning technique for an accurate diagnosis of coronary artery disease**. *Computer methods and programs in biomedicine* 2019, **179**:104992.
43. Khan MU, Aziz S, Naqvi SZH, Rehman A: **Classification of Coronary Artery Diseases using Electrocardiogram Signals**. In: *2020 International Conference on Emerging Trends in Smart Technologies (ICETST): 2020*: IEEE; 2020: 1-5.
44. Ghiasi MM, Zendehboudi S, Mohsenipour AA: **Decision tree-based diagnosis of coronary artery disease: CART model**. *Computer methods and programs in biomedicine* 2020, **192**:105400.
45. Hasan N, Bao Y: **Comparing different feature selection algorithms for cardiovascular disease prediction**. *Health and Technology* 2021, **11**(1):49-62.
46. Reddy KVV, Elamvazuthi I, Aziz AA, Paramasivam S, Chua HN, Pranavanand S: **Heart disease risk prediction using machine learning classifiers with attribute evaluators**. *Applied Sciences* 2021, **11**(18):8352.
47. Zomorodi-moghadam M, Abdar M, Davarzani Z, Zhou X, Pławiak P, Acharya UR: **Hybrid particle swarm optimization for rule discovery in the diagnosis of coronary artery disease**. *Expert Systems* 2021, **38**(1):e12485.
48. Zhang S, Yuan Y, Yao Z, Wang X, Lei Z: **Improvement of the Performance of Models for Predicting Coronary Artery Disease Based on XGBoost Algorithm and Feature Processing Technology**. *Electronics* 2022, **11**(3):315.

**Supplementary Material**

**Supplementary Table 1**. Comparison of our study results with those of previous studies.

| Study | Year | Disease | Eliminated Features | Method | Accuracy (%) |
|---|---|---|---|---|---|
| Abdar et al., [8][29] | 2019 | CAD[1] | Yes | NE-nu-SVC | 94.66 |
| Abdar et al., [9][42] | 2019 | CAD | Yes | N2Genetic-nu-SVM | 93.08 |
| Khan et al. [10][43] | 2020 | CAD | No | SVM-C[2] | 95.60 |
| Ghiasi et al. [11][44] | 2020 | CAD | Yes | CART[3] | 92.41 |
| Hasan and Bao [12][45] | 2021 | CVD[4] | Yes | XGBoost | 73.74 |
| Ghosh et al. [13][33] | 2021 | CVD | Yes | RFBM[5] | **99.05** |
| Reddy et al. [14][46] | 2021 | CVD | No | SMO[6] | 85.14 |
| Reddy et al. [14][46] | 2021 | CVD | Yes | SMO | 86.46 |
| Khan et al. [15] | 2021 | CAD | Yes | SVM-C | 90.26 |
| Zomorodi et al. [16][47] | 2021 | CAD | Yes | Hybrid PSO[7] | 84.25 |
| Singh et al. [17][32] | 2022 | CAD | No | 1 – NELM[8] | **100** |
| Gupta et al. [18][34] | 2022 | CAD | Yes | C-CADZ | **97.37** |
| Zhang et al. [19][48] | 2022 | CAD | Yes | XGBoost | 94.70 |
| **Our study** | **2022** | **CAD** | **No** | **RBFSVM** | **96.00** |

[1] Coronary artery disease (CAD), [2] SVM-Cubic (SVM-C), [3] Classification and regression tree (CART), [4] Cardiovascular Disease (CVD), [5] Random Forest Bagging Method (RFBM), [6] Sequential Minimal Optimization (SMO), [7] Particle Swarm Optimization (PSO), [8] 1-Norm extreme Learning Machine (1 − NELM).